\documentclass[12pt]{iopart}

\usepackage{graphicx,setspace}
\usepackage{dsfont}
\usepackage{iopams}
\usepackage{mathrsfs}
\usepackage{psfrag}
\usepackage{cases}
\usepackage{color}

\newcommand{\p}{{\partial}}

\def\be{\begin{eqnarray}}
\def\ee{\end{eqnarray}}

\begin{document}

\title{Thermal energies of classical and quantum damped oscillators coupled to reservoirs}

\author{T G Philbin and J Anders}
\address{Physics and Astronomy Department, University of Exeter,
Stocker Road, Exeter EX4 4QL, UK.}
\eads{\mailto{t.g.philbin@exeter.ac.uk}}

\begin{abstract}
We consider the global thermal state of classical and quantum harmonic oscillators that interact with a reservoir. Ohmic damping of the oscillator can be exactly treated with a 1D scalar field reservoir, whereas general non-Ohmic damping is conveniently treated with a continuum reservoir of harmonic oscillators. Using the diagonalized Hamiltonian of the total system, we calculate a number of thermodynamic quantities for the damped oscillator: the mean force internal energy, mean force free energy, and another internal energy based on the free-oscillator Hamiltonian. The classical mean force energy is equal to that of a free oscillator, for both Ohmic and non-Ohmic damping and no matter how strong the coupling to the reservoir. In contrast, the quantum mean force energy depends on the details of the damping and diverges for strictly Ohmic damping. These results give additional insight into the steady-state thermodynamics of open systems with arbitrarily strong coupling to a reservoir, complementing results for energies derived within dynamical approaches (e.g.\ master equations) in the weak-coupling regime.
\end{abstract}

\section{Introduction}

Damped oscillators are of importance in numerous experimental and natural settings and their dynamics has been extensively modelled \cite{bre02,wei08}. 
Various approaches to the quantization of such oscillators have been explored. For example, one approach deals solely with the oscillator as the total dynamical system in which case the energy is not conserved~\cite{dek81,um02}. Other approaches add additional degrees of freedom, i.e.\ different kinds of reservoirs, in addition to the system of interest. These reservoir approaches allow the quantization of systems with dissipation to be developed with a time-independent global Hamiltonian, which offers advantages in applying the quantum formalism. The usual practice to arrive at damped motion is to couple the oscillator to additional harmonic oscillators, either a discrete set~\cite{mag59,fey63,cal83,tat87,yu94,wei08} or a continuum~\cite{hut92,phi12,phi13,bar15}.   Alternative models couple the oscillator to a scalar field \cite{unr89}. The quantization techniques employed with reservoirs include path integrals~\cite{fey63,cal83}, canonical quantization~\cite{hut92}, and phenomenological approaches~\cite{yu94}. 

The aim of this paper is to derive thermal-equilibrium energies and free energies of damped oscillators in the classical and quantum regimes. We employ both a scalar field reservoir~\cite{unr89}  and the continuum reservoir~\cite{hut92} , and our approach throughout is based on canonical quantization and diagonalization of the total Hamiltonian. 
Contrary to many master equation approaches which assume an initial state of the oscillator and reservoir that is a product state \cite{cal83,gar85,unr89,wei08,sub12}, we will here assume that the \emph{global} system is in a thermal equilibrium state, implying that it is not in a product state because of the coupling.
We consider two different definitions of the thermal energy of a damped oscillator: the mean-force energy, which includes a contribution from the oscillator-reservoir coupling term in the Hamiltonian~\cite{kir35,jar04,Campisi09}, and an energy based solely on the oscillator part of the Hamiltonian. 
A case of particular interest is Ohmic damping  of the oscillator, where the damping is proportional to velocity, but we also wish to provide expressions for thermal energies for non-Ohmic damping. 
It is well known that the energy of a quantum oscillator with Ohmic damping is divergent~\cite{gra84}, a problem usually treated by the introduction of a cut-off frequency \cite{cal83,gra84,wei08} that changes the damping from strictly Ohmic damping. Interestingly, in the most widely used treatment of damped motion, which begins with a discrete set of harmonic oscillators as a reservoir~\cite{cal83,wei08}, even the classical case contains divergent quantities in the equations of motion. 
Using the scalar-field reservoir \cite{unr89} and the Huttner-Barnett reservoir (a continuum of harmonic oscillators) \cite{hut92} to model Ohmic and non-Ohmic damping, we will obtain finite results for the energies of the oscillator in the classical case, from a dynamics that contains no divergent quantities. We will thereby exhibit within a reservoir treatment that Ohmic damping is physically possible in classical mechanics but impossible in quantum mechanics. A comparison will also be made between the Ohmic-damping thermal energies and those for general damping,  and between classical and quantum thermal energies.

While we here consider the case of a global thermal state, the two reservoir approaches we discuss can readily be applied to investigate non-equilibrium dynamics (see~\cite{unr89}, for example).  The use of reservoirs for non-equilibrium problems is discussed by many authors, e.g.~\cite{cal83,gar85,paz08,alo12,sub12,val15}. 

\section{Ohmic damping with a scalar field reservoir}

Ohmic damping corresponds to the following simple equation of motion of a harmonic oscillator:
\be \label{qdampf}
	\ddot{q} (t) +\gamma \, \dot{q} (t) + \omega_0^2 \, q (t) = f(t),
\ee
where $q$ is the oscillator displacement, $\gamma$ is a damping constant, $\omega_0$ is the frequency of an externally applied potential and $f$ is an external force.  The question now arises as to how a reservoir and coupling can be chosen to obtain (\ref{qdampf}) as the effective equation of motion for the oscillator. 
Coupling the oscillator to a reservoir of harmonic oscillators, either a discrete or continuous set, does not give (\ref{qdampf}) as the equation of motion for $q$, except as a limiting case~\cite{cal83,tat87,wei08} or with an accompanying zero-frequency solution~\cite{phi12}. 
In order to obtain  (\ref{qdampf}) exactly, a reservoir \cite{unr89} consisting of a 1D scalar field $\phi (x, t)$ may be used, linearly coupled to the oscillator. In detail, the Lagrangian is 
\be  \label{lagrangian}
\fl
	L = \frac{1}{2} \left(\dot{q}^{2}(t) - \omega_{0}^{2} \, q^{2} (t) \right)
	+\frac{1}{2} \int_{-\infty}^{\infty} \rmd x \left[\frac{1}{c^2} \, \dot{\phi}^{2} (x, t) - (\p_x \phi (x,t))^2 \right] - \alpha \, \dot{q}(t) \, \phi(0,t),    
\ee
where  $\alpha$ is the coupling constant. 
This gives the equations of motion
\be \label{claseqs}
	\ddot{q} (t) + \omega_0^2 \, q (t) = \alpha \,  \dot{\phi}(0,t) 
	\quad \mbox{and} \quad 
	\frac{1}{c^2} \, \ddot{\phi} (x, t) - \p_x^2\phi (x, t) = - \alpha \, \dot{q}(t) \, \delta(x).
\ee
As was shown previously~\cite{unr89}, the solution for the scalar field is
\be
	\phi(x,t) = -\frac{c}{2} \,  \alpha \, q \left(t- {|x| \over c} \right) + \phi_h(x,t),
\ee
where $\phi_h(x,t)$ is a solution of the homogeneous $\phi$-equation $\ddot{\phi}/c^2-\p_x^2\phi=0$. This yields a $q$-equation of the desired form stated in Eq.~(\ref{qdampf}):
\be  \label{qdampphi}
	\ddot{q} (t) + \gamma \, \dot{q} (t) +\omega_0^2 \, q (t) = \alpha \, \dot{\phi}_h(0,t) 
	\quad \mbox{with} \quad 
	\gamma:=\frac{c}{2} \alpha^2,
\ee
with $\gamma$ proportional to $\alpha^2$.  As a main result of this paper we will now quantize and diagonalise the Hamiltonian for this simple system, which  exhibits exact Ohmic damping.

\section{Quantization and diagonalization of the scalar field model}

From Eq.~(\ref{lagrangian}) one obtains the canonical momenta
\be	\label{canops}
	\Pi_q (t) = \dot{q} (t) - \alpha \, \phi(0,t)
	\quad \mbox{and} \quad 
	\Pi_\phi (x, t) =\frac{1}{c^2} \, \dot{\phi} (x, t).
\ee
We quantize in the Heisenberg picture by imposing the equal-time commutation relations
\be  \label{cancom}
	[\hat{q}(t), \hat{\Pi}_q(t)] = \rmi \hbar
	\quad \mbox{and} \quad 
	[\hat{\phi}(x,t),\hat{\Pi}_{\phi}(x',t)] = \rmi \hbar \, \delta(x-x'),
\ee
while all other commutators of these operators vanish. Here the quantized operators are indicated with hats. Combining Eq.~(\ref{lagrangian}) and Eq.~(\ref{canops}) the quantized Hamiltonian is 
\be  \label{H}
\fl
	\hat{H}= \frac{1}{2} \left[\hat{\Pi}_q^2+ \omega_0^2 \hat{q}^2 \right]
	+\frac{1}{2}\int_{-\infty}^\infty  \rmd x \left[c^2  \hat{\Pi}_{\phi}^2 +(\p_x \hat{\phi})^2\right] 
	+\frac{\alpha}{2}\left[\hat{\Pi}_q \hat{\phi}(0,t) + h.c.\right]
	+\frac{\alpha^2}{2} \hat{\phi}^2(0,t) 
\ee
where $h.c.$ stands for Hermitian conjugate and a Hermitian combination of operators has been taken in the second-last term.

The diagonalization of a Hamiltonian of the general form (\ref{H})
has been described in detail in~\cite{phi12} (see also~\cite{bar15}).  The Hamiltonian (\ref{H}) is diagonalized by transforming it into the normal form
\be	\label{HC} 		
	\hat{H} = \frac{1}{2} \int_{-\infty}^\infty \rmd k \, 
		\hbar\omega \, \left[\hat{C}^\dagger(k)\hat{C}(k)+\hat{C}(k)\hat{C}^\dagger(k)\right] 		\quad \mbox{with} \quad \omega:=c|k| ,
\ee
where $\hat{C}^\dagger(k)$ and $\hat{C}(k)$ are the creation and annihilation operators for a free scalar field, 
\be  \label{Psi}
	\hat{\Psi}(x,t)=\int_{-\infty}^\infty \rmd k \, \sqrt{\frac{c^2\hbar}{4\pi\omega}}\left[  \hat{C}(k)e^{\rmi kx-\rmi\omega t}  + \hat{C}^\dagger(k)e^{-\rmi kx+\rmi\omega t} \right].
\ee
The ladder operators obey the standard bosonic field commutation relations
\be
{[} \hat{C}(k),\hat{C}^\dagger(k')]=\delta(k-k') \quad \mbox{and} \quad [\hat{C}(k),\hat{C}(k')]=0.  \label{CCd}
\ee
The diagonalization also goes through classically, with commutators replaced by Poisson brackets.
We show in the Appendix that the relationship between the dynamical variables appearing in Eq.~(\ref{H}) and in Eq.~(\ref{HC}) is given by 
\be
	\hat{q}(t) = \int_{-\infty}^\infty \rmd k \, \sqrt{\frac{\hbar\omega}{4\pi}} \left[ \frac{\rmi \, c \, \alpha}{\omega^2 - \omega^2_0 + \rmi \gamma \omega} \, \hat{C}(k) \, e^{-\rmi\omega t}  +\mathrm{h.c.} \right],  \label{qC1} \\
\fl
\hat{\phi}(x,t)  = \int_{-\infty}^\infty \rmd k  \, \sqrt{\frac{c^2 \hbar}{4\pi \omega}} \left[ 
\left( e^{\rmi k x} -  \frac{\rmi \, \gamma \, \omega}{\omega^2 - \omega^2_0 + \rmi  \gamma \omega} \,  e^{\rmi \omega \, |x|/c}   \right) \, \hat{C}(k) \, e^{-\rmi\omega t} +\mathrm{h.c.} 
\right],     \label{qC2}  \\
 \hat{\Pi}_q(t)  = \dot{\hat{q}}(t) - \alpha \, \hat{\phi}(0,t),  \label{qC3}  \\
   \hat{\Pi}_\phi(x,t) =\frac{1}{c^2} \, \dot{\hat{\phi}}(x,t).   \label{qC4}
\ee

\section{Thermal equilibrium of the scalar-field model}  \label{sec:thermal}

We are now ready to calculate thermal equilibrium expectation values of the total system and deduce the internal energy and free energy of the open quantum oscillator, as well as of its classical counterpart. A common assumption in dynamical approaches is that the reservoir is always in a thermal state while the system evolves in time towards a steady state~\cite{bre02,wei08}. In contrast, we here consider the case where the oscillator and system have been interacting for so long that a global thermal state has been reached. 

  For a non-equilibrium treatment of the scalar-field reservoir, in the context of the quantum-classical transition, see~\cite{unr89}. In global thermal equilibrium, i.e. $\rho_{\mathrm{tot}} (\beta) = {e^{-\beta \hat{H}} \over Z}$, $\beta^{-1}=k_BT$, the normal modes of the total system have the  expectation values:
\be
\fl
	\langle\hat{C}^\dagger(k) \hat{C}(k')\rangle_{\mathrm{tot}} &= \mathcal{N}(\omega) \, \delta(k-k') \label{CCdthermal} 
	\quad \mbox{with} \quad 
	\mathcal{N}(\omega)=\left[\exp\left(\frac{\hbar\omega}{k_BT}\right)-1\right]^{-1} \label{planck}   \\
	\fl
\langle\hat{C}(k)\hat{C}(k')\rangle_{\mathrm{tot}} & =0.  \label{CCthermal}
\ee
These equations now determine the thermal correlation functions of the $q$-oscillator. 
Using Eq.~(\ref{qC1}) it is straightforward to show that
\be  \label{qqtime}
\fl
	\frac{1}{2}\left\langle \hat{q}(t)\hat{q}(t') + h.c. \right\rangle_{\mathrm{tot}}=\frac{\hbar}{\pi}\int_{0}^\infty \rmd \omega  \frac{ \gamma \omega}{(\omega^2-\omega^2_0)^2+\gamma^2\omega^2} \cos\left[\omega(t-t')\right]\coth\left(\frac{\hbar\omega}{2k_BT}\right).
\ee
This result can also be obtained by using Eq.~(\ref{qdampf}) and appealing to the fluctuation-dissipation theorem~\cite{gra84}. In this scalar field model the fluctuation-dissipation theorem does not have to be imposed, rather it arises as a consequence of thermal equilibrium of the total system as formalised in Eqs.~(\ref{CCdthermal})--(\ref{CCthermal}). The above autocorrelation function is finite and depends on the damping parameter $\gamma$ but further results show the quantum case to be unphysical (see~\cite{gra84} and below).

To calculate the thermodynamic quantities for the oscillator, such as its internal energy, one must quantify the contributions that come from its coupling to the reservoir which causes the damping~\cite{GT09}. The energy of an open oscillator, which is part of a total thermal state, can be accounted for by the Hamiltonian of mean force~\cite{kir35}. The Hamiltonian of mean force appears naturally in non-equilibrium work relations~\cite{jar04,Campisi09,phi14}, in the thermodynamic analysis of the second law and Landauer's principle for a damped oscillator~\cite{HSAL11}, in the evolution to steady state of open systems coupled to a thermal reservior~\cite{sub12}, and also in the Casimir effect~\cite{phi13}.
The Hamiltonian of mean force for the oscillator is defined as \cite{Campisi09}
\be  \label{defH*}
	\hat{H}^\star (\beta)  = - {1 \over \beta} \, \ln {\tr_{\phi} [e^{-\beta \hat{H}}] \over Z_\phi}, \qquad Z_\phi=\tr_{\phi} [e^{-\beta \hat{H}_{\phi}}], \qquad \beta=\frac{1}{k_B T},
\ee
where the traces are taken over the scalar field $\phi$ only and $\hat{H}_{\phi} = \frac{1}{2}\int_{-\infty}^\infty  \rmd x \left[c^2  \hat{\Pi}_{\phi}^2 +(\p_x \hat{\phi})^2\right]$ is the free scalar-field Hamiltonian, a part of the total Hamiltonian $\hat{H}$ given in (\ref{H}). Thus, $Z_\phi$ in (\ref{defH*}) is the partition function of a free scalar field in thermal equilibrium. The partition function $Z^\star$ associated with $\hat{H}^{\star}$ is 
\be   \label{Zstar}
	Z^\star=\mathrm{tr}_{S}\left[e^{-\beta\hat{H}^\star}\right]
	=\frac{Z}{Z_\phi}, \qquad Z=\mathrm{tr}[e^{-\beta\hat{H}}],
\ee
where $Z$ is the partition function of the total system. The free energy $F^\star$ associated with $Z^\star$ is, from (\ref{Zstar}),
\begin{equation}
F^{\star}=-\beta^{-1}\ln(Z^{\star})=F-F_{\phi},  \label{free-energy}
\end{equation}
where $F=-\beta^{-1}\ln(Z)$ is the free energy of the total system and $F_{\phi}=-\beta^{-1}\ln(Z_{\phi})$ is the free energy of a free scalar field in thermal equilibrium. From the standard thermodynamic relation $F=U-TS$ relating the free energy to the internal energy $U$ and entropy $S$, we can find the internal energy associated with the mean force free energy $F^\star$. From (\ref{free-energy}) we have
\begin{equation}  \label{FstarUS}
F^{\star}=U-U_\phi-T(S-S_\phi),
\end{equation}
where $U$ is the total internal energy, $U_\phi$ is the internal energy of a free scalar field, $S$ is the total entropy and $S_\phi$ is the entropy of a free scalar field. Equation (\ref{FstarUS}) shows that the mean force thermal energy $U^\star$ associated with $F^\star$ is
\begin{equation}  \label{U*formula}
U^\star=U-U_\phi=\langle\hat{H}\rangle_{\mathrm{tot}}-Z_\phi^{-1}\tr_{\phi} [\hat{H}_{\phi} e^{-\beta \hat{H}_{\phi}}],
\end{equation}
which is the total thermal energy minus the thermal energy of a free scalar field. Note that $U^\star$ is \emph{not} defined as the expectation value of $\hat{H}^\star$ in the global thermal state, i.e. $U^\star :\neq  \langle \hat{H}^\star \rangle_{\mathrm{tot}}$. The total Hamiltonian (\ref{H}) can be  rewritten with (\ref{qC3}) in terms of $\hat{q}$ and $\hat{\phi}$  as
\be 	\label{energyop}
	\hat{H}=\frac{1}{2} \, \dot{\hat{q}}^2+\frac{1}{2}\omega_0^2 \, \hat{q}^2
	+\frac{1}{2}\int_{-\infty}^\infty  \rmd x \left[\frac{1}{c^2} \, \dot{\hat{\phi}}^2+(\p_x \hat{\phi})^2\right].
\ee
Now $U^\star$ can be obtained as the expectation value of (\ref{energyop}) in the global thermal state $\rho_\mathrm{tot}$ calculated using (\ref{qC1}), (\ref{qC2}), (\ref{CCdthermal}) and (\ref{CCthermal}), and dropping all $\gamma$-independent terms in the final scalar-field contribution (this subtracts out the free scalar-field energy as required by (\ref{U*formula})). Remarkably, the $\gamma$-dependent terms arising from the scalar-field part of (\ref{energyop}) cancel out in the expectation value, and so $U^\star$ is the same as would be obtained from just the $q$-terms in (\ref{energyop}). The result for $U^\star$ is
\be  \label{U*}
	U^\star=\frac{\hbar}{2\pi}\int_{0}^\infty \rmd \omega \,  
	\frac{ \gamma \omega \, (\omega^2+\omega^2_0)}{(\omega^2-\omega^2_0)^2+\gamma^2\omega^2} \, 
	\coth\left(\frac{\hbar\omega}{2k_BT}\right).
\ee
This expression diverges for any temperature $T$ which proves that Ohmic damping of a quantum oscillator is unphysical.  Such divergences are often avoided by introducing a  high frequency cut-off at the outset \cite{gra84} which in turn results in approximately Ohmic damping for those frequencies that can be supported by the system. 

\subsection{Classical Ohmic damping}

Classically the $\hat{C}(k)$ in (\ref{Psi}) become the complex amplitudes $C(k)$ of the normal modes, and the thermal-equilibrium expectation value corresponding to (\ref{CCdthermal}) is $\langle C^*(k) \, C(k')\rangle=\delta(k-k') \, k_B\, T /(\hbar\omega)$. The occurrence of $\hbar$ in this classical expression is due to the $\hbar$-dependent normalization of the complex amplitudes in (\ref{Psi}). We will derive classical thermal results as limits $\hbar \to 0$ of the quantum expressions, but they may of course be directly obtained from the classical normal-mode expectation values.

In the classical limit of (\ref{U*}) the  mean force internal energy is finite and neatly gives
\be  \label{eq:clU*}
	U^\star= {k_B \, T \over \pi} \, \int_{0}^\infty \rmd \omega \,  \frac{ \gamma  \, (\omega^2+\omega^2_0)}{(\omega^2-\omega^2_0)^2+\gamma^2\omega^2} 
	= k_B \, T,
\ee
which is the same result as for an {\it undamped} oscillator. This result, $U^\star=k_B \, T$, is actually true not only for Ohmic damping, but also for very general damping, see next section.  

An alternative definition of the internal energy of the  oscillator that is often considered~\cite{GT09} is 
\be \label{eq:U}
	U := {1 \over 2} \, \langle \, \dot{q}^2 + \omega_0^2 \, q^2 \rangle_{\mathrm{tot}},
\ee
where the expectation value is again taken in the global thermal state. Note that this energy could depend on the coupling to the reservoir through both the correlations in the thermal state and the dependence of $q$ and $\dot{q}$ on the coupling $\alpha$. As discussed after Eq.~(\ref{energyop}), for Ohmic damping it turns out that the scalar field contribution to $U^\star$ cancels out, i.e. the mean force energy is the \emph{same} as would be obtained using the definition of $U$ in Eq.~(\ref{eq:U}). Thus for the case of Ohmic damping one obtains $U = U^\star = k_B \, T$. Interestingly, for general damping $U^\star$ differs from $U$, see next section.

\section{Huttner-Barnett reservoir  and general damping}

In the previous sections the reservoir was taken to be a 1D scalar field because this leads very simply to Ohmic damping of the $q$-oscillator. For general damping, a reservoir of harmonic oscillators can be used.  In~\cite{phi12,phi13,bar15} an oscillator coupled to a reservoir consisting of a continuum of harmonic oscillators $\{X_\omega:\omega \in [0, \infty)\}$ was considered. This continuum reservoir was introduced by Huttner and Barnett~\cite{hut92}. The resulting Hamiltonian is \cite{phi12} 
\be  \label{Hhb}
\fl
	\hat{H}=\frac{1}{2} \, \hat{\Pi}_q^2+\frac{1}{2} \, \omega_0^2 \, \hat{q}^2
	+\frac{1}{2}\int_0^\infty\rmd\omega\left(\hat{\Pi}_{X_{\omega}}^2+\omega^2\hat{X}_\omega^2\right) 
	-\frac{1}{2}\int_0^\infty\rmd\omega\,\alpha(\omega)\left[\hat{q} \, \hat{X}_\omega+\hat{X}_\omega \, \hat{q}\right], 
\ee
where $\alpha (\omega)$ is a function describing the coupling between the oscillator and reservoir modes. 
The resulting dynamics of the $q$-oscillator is governed by a complex susceptibility $\chi(\omega)=\chi^\star(-\omega)$ whose  imaginary part is proportional to $\alpha^2(\omega)$. The susceptibility obeys Kramers-Kronig relations and so $\chi(\omega)$ is analytic in the upper-half complex $\omega$-plane~\cite{phi12}. In addition there is a sufficient condition on $\chi (\omega)$ for the total Hamiltonian to be diagonalizable, namely~\cite{phi12}
\be \label{ineqgen}
 	\int_0^\infty \rmd \omega\, \mathrm{Im}[\chi(\omega)] <\frac{\pi}{2}.
\ee
The damping term in the effective equation of motion for $q$, see (\ref{qdampf}) for the Ohmic damping case, now features a damping kernel, $\omega_0^2\int_{-\infty}^\infty \rmd t'\,\chi(t-t') \, \hat{q}(t')$. The position operator for the $q$-oscillator then becomes~\cite{phi12}
\be \label{eq:gendamp}
	\hat{q}(t)=\int_0^\infty\rmd\omega \sqrt{\frac{\hbar}{2\omega}} \, \left[ \frac{-\alpha(\omega)}{\omega^2-\omega_0^2[1-\chi(\omega)]} \, \hat{C}(\omega) \, e^{-\rmi\omega t} +\mathrm{h.c.}\right], 
\ee
where $\hat{C}(\omega)$ are the annihilation operators for the modes that diagonalize the full Hamiltonian (\ref{Hhb}), analogous to  Eq.~(\ref{HC}). 
The explicit form of the Hamiltonian in (\ref{Hhb}) allows the calculation of the mean force internal energy~\cite{phi12}:
\be \label{Hq}
	U^\star= \frac{\hbar}{2\pi} \, \int_{0}^\infty \rmd\omega  \coth\left(\frac{\hbar\omega}{2k_BT}\right) 
 	 \mathrm{Im}\left\{\frac{\omega_0^2 \left[\omega\, {d\chi(\omega) \over d\omega} - \chi(\omega)+1 \right]+\omega^2}{\omega_0^2 \left[1-\chi(\omega)\right]-\omega^2}\right\}. 
\ee
It is easy to obtain from the results in~\cite{phi12} that the internal energy $U$ defined by (\ref{eq:U}) differs from (\ref{Hq}) by not having the $\chi$-dependent terms in the numerator inside the curly brackets, i.e.
\be \label{Ugen}
	U= \frac{\hbar}{2\pi} \, \int_{0}^\infty \rmd\omega  \coth\left(\frac{\hbar\omega}{2k_BT}\right) 
 	 \mathrm{Im}\left\{\frac{\omega_0^2 +\omega^2}{\omega_0^2 \left[1-\chi(\omega)\right]-\omega^2}\right\}. 
\ee
This means that for general damping the two energy expressions (\ref{Hq}) and (\ref{Ugen}) will differ (i.e. $U^*\neq U$) both in the quantum and classical cases. For the classical oscillator, both $U^\star$ and $U$ can be evaluated exactly for arbitrary $\chi$, see below.

\subsection{Ohmic damping}

As can be seen by comparing Eq.~(\ref{Hq}) with the Ohmic expression Eq.~(\ref{U*}), Ohmic damping corresponds to choosing a ``susceptibility"
\be \label{chiOh}
	\chi(\omega)=\frac{\rmi \gamma \omega}{\omega_0^2}.
\ee
However, this choice is not physical as it obeys neither Kramers-Kronig relations nor condition (\ref{ineqgen}), reflecting the well-known result that strictly Ohmic damping cannot be treated by a reservoir of harmonic oscillators \cite{tat87}. (We note, however, that the case of Ohmic damping when accompanied by a zero-frequency solution for the $q$-oscillator \emph{can} be properly treated by a valid susceptibility~\cite{phi12}.)

\medskip

Despite the fact that the Ohmic damping ``susceptibility'' (\ref{chiOh}) is not a valid choice, inserting it into the general result (\ref{Hq}) gives the Ohmic-damping result  (\ref{U*}) derived with the scalar field reservoir. Also, in (\ref{Hq}) the $\chi$-dependent terms in the numerator inside the curly brackets cancel out for this choice (\ref{chiOh}), which again gives $U^\star=U$ in the Ohmic damping case, as found in the previous section.

\subsection{Classical limit}

We now derive the important result that in the classical limit $\hbar \to 0$ one obtains $U^\star=k_B \, T$ for almost any susceptibility, whereas $U$ depends on the susceptibility (i.e.\ the damping).
If we take the Im outside the integration in (\ref{Hq}) then the real part of the resulting integral does not converge. However, setting the lower integration limit to $-\infty$ and dividing the whole integral by 2 removes the diverging real part, because the real part of the integrand is odd in $\omega$ (recall that $\chi(\omega)=\chi^*(-\omega)$, so  $\mathrm{Re}[\chi(\omega)]$ is even and $\mathrm{Im}[\chi(\omega)]$ is odd).  The new integral from $-\infty$ to $+\infty$ now requires the pole at $\omega=0$ to be treated as a principal value so that the integration picks out the even part of the integrand. The resulting expression in the classical limit is
\be  \label{U*clasinet}
	U^\star= \frac{k_BT}{2\pi} \, \mathrm{Im} \left[  \mathrm{P}\int_{-\infty}^\infty \rmd\omega  \, \frac{\omega_0^2\left[\omega\, {d\chi(\omega) \over d\omega} -\chi(\omega)+1\right]+\omega^2}{\omega(\omega_0^2\left[1-\chi(\omega)\right]-\omega^2)} \right]
	=k_BT, 
\ee
where P denotes principal value. For very general $\chi(\omega)$, the principal-value integral in (\ref{U*clasinet}) evaluates to $2\pi\rmi$, as shown by the following analysis. If $\omega\, {d\chi(\omega) \over d\omega}|_{\omega=0}=0$, then the integrand in (\ref{U*clasinet}) has a simple pole at $\omega=0$. Consider the same integrand integrated over a closed contour $C$ in the complex $\omega$-plane that runs along the real line but goes below the pole at $\omega=0$, and then closes anti-clockwise in a large semicircle in the upper-half plane. This contour integral can be decomposed as
\be \label{ints}
\oint_C dz =  \mathrm{P}\int_{-\infty}^\infty \rmd\omega +\int_{C_\epsilon} dz +  \int_R dz,
\ee
where $C_\epsilon$ is an infinitesimal semicircle running anti-clockwise around the pole at $\omega=0$ and $R$ is a large semicircle of radius $R$ in the upper-half plane taken in the anti-clockwise direction. As the integrand in (\ref{U*clasinet}) is analytic everywhere inside $C$ except at the simple pole at $\omega=0$ (recall that $\chi(\omega)$ is analytic in the upper-half plane), the integral around $C$ is $2\pi\rmi$. The integral along $C_\epsilon$ is $\pi\rmi$, and it is easy to show that the integral along the semicircle of radius $R$, as $R\to\infty$, is $-\pi\rmi$. From (\ref{ints}) this shows that the principal-value integral in (\ref{U*clasinet}) is $2\pi\rmi$ so we obtain the classical result $U^\star=k_BT$ for very general susceptibility.

The classical limit of the energy $U$ (\ref{Ugen}) is
\be
	U= \frac{k_BT}{2\pi} \mathrm{Im}\left[  \mathrm{P}\int_{-\infty}^\infty \rmd\omega  \, \frac{\omega_0^2 +\omega^2}{\omega(\omega_0^2\left[1-\chi(\omega)\right]-\omega^2)} \right]. \label{Ugenclas}
\ee
In evaluating this integral the only changes to the above analysis of $U^\star$ are that the integral around the closed contour $C$ is now $2\pi\rmi/[1-\chi(0)]$, and the integral along $C_\epsilon$ is $\pi\rmi/[1-\chi(0)]$, assuming $\chi(0)$ is finite and not equal to 1. Hence, the classical internal energy $U$ becomes 
\be  \label{Uclas}
U=k_BT \left[1+\frac{\chi(0)}{2[1-\chi(0)]} \right], \qquad \chi(0)\neq1.
\ee
This depends on the damping unless $\chi(0)=0$. The general result (\ref{Uclas}) reproduces the classical value $U=k_BT$ for Ohmic damping obtained in the last section if we again substitute the ``susceptibility" (\ref{chiOh}), because in this case  $\chi(0)=0$. Thus, in the classical limit considered here, the general damping dependence of the energy $U$ (\ref{Uclas}) contrasts with the damping independence of the mean force energy $U^\star =k_B \, T$.

\subsection{Comparison of energies for different damping}

Table \ref{tab:comparison} summarises the classical and quantum results for $U$ and $U^\star$ with different types of damping. The calculations above showed that for Ohmic damping of a classical oscillator, the energy can be taken to be either $U$ or $U^\star$ as both reduce to $k_B T$. In contrast, for a quantum oscillator, both $U$ and $U^\star$ diverge for Ohmic damping. For general non-Ohmic damping, classically $U^\star$ is always $k_B \, T$ whereas $U$ depends on the damping. Quantum mechanically both $U^\star$ and $U$ depend on the damping, however they are not equal. Note that for a classical oscillator small deviations from Ohmic damping result in small changes in $U$. In contrast, for a quantum oscillator with Ohmic damping the energies diverge, while small deviations from Ohmic damping make $U$ and $U^\star$ finite. Thus, even for oscillators that are classically well described by Ohmic damping,  their quantum (including zero-point) energies are entirely determined by the deviations from Ohmic damping.

\begin{table}[ht]
\begin{center}
\begin{tabular}{|l|c|c|}
\hline
damping & classical & quantum \\
\hline
no damping & $U = U^\star = k_B \, T$ & $U = U^\star = {\hbar  \omega_0 \over 2} \, \coth {\hbar \beta \omega_0 \over 2}$ \\ 
Ohmic damping & $U = U^\star = k_B \, T$  & $U, U^\star \to \infty$ \\ 
general damping & $U = k_B \, T \left[1+\frac{\chi(0)}{2[1-\chi(0)]} \right]  \neq U^\star = k_B \, T$  & $U  \neq U^\star$, see (\ref{Ugen}) and (\ref{Hq})  \\ 
\hline
\end{tabular}
\caption{\label{tab:comparison}
Table comparing energies for classical and quantum damped oscillators, for no damping, Ohmic damping, and general non-Ohmic damping described by a susceptibility $\chi (\omega)$.
 }
\end{center}
\end{table}

\subsection{Free energy and entropy}

Finally, it is also interesting  to derive the Helmholtz free energy $F^\star$ arising from the mean force energy $U^\star$ for general damped oscillators.  Using the standard thermodynamic relations $F^\star = U^\star-TS^\star$ and $S^\star= - {\p F^\star \over \p T}$,  we obtain $U^\star =-T^2 {d \over d T} {F^\star \over T}$, which gives $F^\star$ as
\be \label{FU}
F^\star= -T \int \rmd T \, \frac{U^\star}{T^2} + aT,
\ee
for some constant $a$. The entropy $S^\star= - {\p F^\star \over \p T}$ and must vanish at $T=0$ in line with the third law of thermodynamics, and this allows the value of $a$ in (\ref{FU}) to be determined. The results for $F^\star$ and $S^\star$ are
\be
\fl
F^*=& \frac{k_BT}{\pi}\int_{0}^\infty \rmd\omega   \ln\left[\sinh\left(\frac{\hbar\omega}{2k_BT}\right)\right]  
 	 \mathrm{Im}\left\{\frac{\omega_0^2\left[\omega\, {d \chi(\omega) \over d\omega}-\chi(\omega)+1\right]+\omega^2}{\omega(\omega_0^2\left[1-\chi(\omega)\right]-\omega^2)}\right\} +k_BT \ln2,  \nonumber \\
	 \fl
 \label{F*quan}   \\
	 \fl
S^\star = & \frac{\hbar}{2\pi}\int_{0}^\infty \rmd\omega  \left\{   \frac{1}{T} \coth\left(\frac{\hbar\omega}{2k_BT}\right)-  \frac{2k_B}{\hbar\omega} \ln\left[\sinh\left(\frac{\hbar\omega}{2k_BT}\right)\right]   \right\} \nonumber \\
\fl
& \qquad\qquad\quad	\times \mathrm{Im}\left\{\frac{\omega_0^2\left[\omega\, {d\chi(\omega) \over d\omega}-\chi(\omega)+1\right]+\omega^2}{\omega_0^2\left[1-\chi(\omega)\right]-\omega^2}\right\}  -k_B \ln2.  \label{S*quan}
\ee
To verify that the entropy (\ref{S*quan}) vanishes at $T=0$, first note that the $T$-dependent factor in the integral reduces to ${2 k_B \over \hbar\omega} \ln 2 $ in the $T\to 0$ limit. The integral is then proportional to an integral evaluated above, see (\ref{U*clasinet}).

The classical limit of the free energy (\ref{F*quan}) is
\be  \label{F*clas}
\fl
F^\star= \frac{k_BT}{\pi}\int_{0}^\infty \rmd\omega  \ln\left(\frac{\hbar \omega}{2k_BT}\right)
 	 \mathrm{Im}\left\{\frac{\omega_0^2\left[\omega\,{d\chi(\omega) \over d\omega}-\chi(\omega)+1\right]+\omega^2}{\omega_0^2\left[1-\chi(\omega)\right]-\omega^2}\right\} 
	 +k_BT \ln2,
\ee
where $\hbar$ still appears as the phase space volume element, which will cancel in free energy differences \cite{LL}.  In contrast to the damping independence of $U^\star$, the classical free energy $F^\star$ does depend on the details of the damping in almost all cases.  The notable exception is Ohmic damping, for which $F^\star$  can be found with the ``susceptibility" (\ref{chiOh}). The resulting integral can be evaluated exactly and gives the free energy of an undamped oscillator, i.e. $F^\star=k_B \, T\ln \left( {\hbar \omega_0 \over k_B \, T} \right)$, independent of $\gamma$. Alternatively, the classical Ohmic-damping result can be evaluated using the scalar-field reservoir of the last section by the same analysis leading from (\ref{Hq}) to (\ref{F*clas}).

\section{Conclusions}

To model damped harmonic motion we considered two time-independent Hamiltonians for an oscillator coupled to a reservoir, one with a scalar-field reservoir and one with a Huttner-Barnett reservoir. Using the diagonalised Hamiltonians we derived expressions for thermodynamic quantities of  a damped oscillator, for both Ohmic and general damping, when the total system is in a global thermal state. These were evaluated for both the quantum and classical regimes. We recovered the fact that  strictly Ohmic damping of a \emph{quantum} oscillator cannot physically occur due to divergences forcing one to abandon the exact Ohmic regime. In contrast, Ohmic damping of a \emph{classical} oscillator can be treated exactly using the scalar-field reservoir, giving finite and physically meaningful results. The diagonalized form of the Hamiltonians allowed the calculation of thermal energies  of the oscillator, i.e. the mean force energy $U^*$ and its corresponding free energy $F^\star$, and the commonly used internal energy $U$. We found that classically $U^\star=k_B T$ for any damping type, no matter how strong the coupling to the reservoir. This demonstrates a remarkable and non-trivial property of the classical mean force energy.  In contrast to $U^\star$, the classical internal energy $U$ and the mean force free energy  $F^\star$ do depend on the coupling strength for general non-Ohmic damping. 

For \emph{quantum} oscillators it is surprising that while strictly Ohmic damping is plagued with divergences, infinitesimal changes to the damping result in finite expressions for both $U^\star$ and $U$. In addition, the quantum mean force energy $U^\star$ does depend on the coupling, as do $U$ and $F^\star$.
The treatment of classical and quantum open systems often assumes initial product states between the system of interest and a reservoir that are then evolved with a global Hamiltonian, and under a number of assumptions, to a long-time steady-state~\cite{hat01,bre02,wei08}. In contrast we here considered the stationary situation of the total system being in a global thermal state. The results presented add a new perspective on the thermodynamics of open systems with arbitrarily strong coupling to a reservoir. For example, the different energy measures may be of significance when calculating efficiencies of small scale and quantum engines that operate between equilibrium configurations in the strong coupling limit~\cite{kos}. Such engine cycles may show departures from standard thermodynamics which assumes weak coupling. Finally, extending the thermal equilibrium analysis of the Huttner-Barnett reservoir presented here to analyse the non-equilibrium dynamics of damped oscillators is an interesting topic for future investigation.

\section*{Acknowledgements}
We thank I. Hooper and S. Horsley for discussions that led to this work. JA acknowledges support by EPSRC (EP/M009165/1).


\appendix
\section*{Appendix}
\setcounter{section}{1}
Here we describe the diagonalization of the Hamiltonian (\ref{H}). The procedure is very similar to the diagonalization of the damped harmonic oscillator with a reservoir composed of a continuum of harmonic oscillators~\cite{phi12,bar15} (which is in turn similar to part of the Huttner-Barnett model~\cite{hut92}). 

We seek a linear transformation between the dynamical variables in (\ref{H}) and (\ref{HC}). It is more convenient to work with the time-dependent operators
\begin{eqnarray}
 \hat{C}(k,t)= \hat{C}(k)e^{-\rmi\omega t}, \qquad  \hat{C}^\dagger(k,t)= \hat{C}^\dagger(k)e^{\rmi\omega t},    \label{Ct} \\
 {[} \hat{C}(k,t), \hat{H} ] = \hbar \omega \hat{C}(k,t),  \label{CHC}
\end{eqnarray}
the last relation following from (\ref{HC}) and (\ref{CCd}). The required transformation must take the form
\begin{eqnarray}
\fl
\hat{q}(t)=\int_{-\infty}^\infty \rmd k  \left[  f_q(k) \hat{C}(k,t)  +\mathrm{h.c.} \right],  \quad  \hat{\Pi}_q(t)=\int_{-\infty}^\infty \rmd k  \left[  f_{\Pi_q}(k) \hat{C}(k,t)  +\mathrm{h.c.} \right]   \label{trans1} \\
\hat{\phi}(x,t)=\int_{-\infty}^\infty \rmd k  \left[  f_\phi(x,k) \hat{C}(k,t)  +\mathrm{h.c.} \right], \label{trans2} \\
 \hat{\Pi}_\phi(x,t)=\int_{-\infty}^\infty \rmd k  \left[  f_{\Pi_\phi}(x,k) \hat{C}(k,t)  +\mathrm{h.c.} \right],   \label{trans3}
\end{eqnarray}
for some unknown functions $ f_q(k)$, etc. The commutation relations (\ref{CCd}) with (\ref{trans1})--(\ref{trans3}) give:
\begin{eqnarray}
f_q(k)=[\hat{q}(t),\hat{C}^\dagger(k,t)],     \qquad  f_{\Pi_q}(k)=[\hat{\Pi}_q(t),\hat{C}^\dagger(k,t)],  \label{fC1} \\
f_\phi(x,k)=[\hat{\phi}(x,t),\hat{C}^\dagger(k,t)],     \qquad  f_{\Pi_\phi}(x,k) = [\hat{\Pi}_\phi(x,t),\hat{C}^\dagger(k,t)].  \label{fC2}
\end{eqnarray}
The transformation (\ref{trans1})--(\ref{trans3}) must be invertible, which together with (\ref{cancom}), (\ref{fC1}) and (\ref{fC2}) implies
\begin{eqnarray}
\fl
\hat{C}(k,t)=-\frac{\rmi}{\hbar}{\Bigg\{} & f^*_{\Pi_q}(k)\hat{q}(t)-f^*_q(k)\hat{\Pi}_q(t)    \nonumber \\
\fl
 &  \left.    +\int_{-\infty}^\infty\rmd x \left[f^*_{\Pi_\phi}(x,k)\hat{\phi}(x,t)-f^*_\phi(x,k)\hat{\Pi}_{\phi}(x,t)\right]\right\}.   \label{Cfq}
\end{eqnarray}
We find equations for the $f$-coefficients in (\ref{trans1})--(\ref{trans3}) as follows. Insert (\ref{Cfq}) and (\ref{H}) into (\ref{CHC}) and simplify using (\ref{cancom}). This gives an expression for $\hat{C}(k,t)$ which can be compared with (\ref{Cfq}) to find
\begin{eqnarray}
\fl
f_{\Pi_q}(k)+\alpha f_\phi(0,k)=-\rmi\omega f_q(k), \quad  \rmi\omega  f_{\Pi_q}(k)=\omega_0^2 f_q(\omega),  \label{fPifq} \quad
c^2f_{\Pi_\phi}(x,k)=-\rmi\omega f_\phi(x,k), \\
  \rmi\omega f_{\Pi_\phi}(x,k)=  \alpha f_{\Pi_q}(k)\delta(x)-\p_x^2 f_\phi(x,k) + \alpha^2 f_\phi(0,k)\delta(x)  ,   \label{fPifX}
\end{eqnarray}
which give
\be
\fl
\omega_0^2 f_q(k)  =   \omega^2 f_q(k)- \rmi \alpha \omega f_\phi(0,k), \quad 
\frac{\omega^2}{c^2} f_\phi(x,k)  = - \rmi \alpha \omega f_q(k) \delta(x) - \p_x^2 f_\phi(x,k) .  \label{feqs}
\ee
These are the same as the classical equations (\ref{claseqs}) in the frequency domain and their solution is
\be  \label{fsol}
\fl
 f_\phi(x,k) = -\frac{1}{2}c\alpha  e^{\rmi \omega |x|/c}  f_q(k) + h_\phi(k) e^{\rmi k x}, \quad 
  f_q(k)=\frac{\rmi  \alpha \omega h_\phi(k)}{\omega^2-\omega^2_0+\rmi\gamma\omega} ,
\ee
where $h_\phi(k)$ is the amplitude of the solution to the homogeneous $f_\phi$ equation ($\alpha=0$).

The value of $h_\phi(k)$ is determined by the fact that (\ref{Cfq}) and its Hermitian conjugate have commutator $[\hat{C}(k,t),\hat{C}^\dagger(k',t)]=\delta(k-k')$ (see (\ref{CCd}) and (\ref{Ct})). A tedious calculation shows that this commutator holds with (\ref{Cfq}) expanded in the solutions for the $f$-coefficients if
\be  \label{hphi}
h_\phi(k)= \sqrt{\frac{c^2 \hbar}{4\pi \omega}}.
\ee
The commutator $[\hat{C}(k,t),\hat{C}(k',t)]=0$ is identically satisfied by (\ref{Cfq}) with the solutions for the $f$-coefficients. The expansions (\ref{trans1})--(\ref{trans3}) have now been determined and give (\ref{qC1})--(\ref{qC3}). Consistency of the diagonalization is demonstrated by showing that (\ref{qC1})--(\ref{qC3}) obey the commutation relations (\ref{cancom}) because of (\ref{CCd}).

\end{document}